\documentclass[prl,twocolumn,showpacs,preprintnumbers,amsmath,amssymb]{revtex4}
\usepackage{graphicx}
\usepackage{dcolumn}% Align table columns on decimal point
\usepackage{bm}% bold math
\usepackage{color}
\begin{document}
\title{Temperature dependent transport in suspended graphene}
\author{K. I. Bolotin$^1$}
\author{K. J. Sikes$^2$}
\author{J. Hone$^3$}
\author{H. L. Stormer$^{1,2,4}$}
\author{P. Kim$^1$}
\affiliation{Depts. of $^1$Physics, $^2$Applied Physics, $^3$Mechanical Engineering, Columbia University, New York, NY 10027}
\affiliation{$^4$Bell Labs, Alcatel-Lucent Technologies, Murray Hill, NJ 07974}
\date{\today}

\begin{abstract}
 The resistivity of ultra-clean suspended graphene is strongly temperature ($T$) dependent for $5$~K$< T < 240$~K. At $T\sim5$~K transport is near-ballistic in a device of $\sim2~\mu$m dimension and a mobility $\sim 170,000$~cm$^2$/Vs. At large carrier density, $n>0.5\times$10$^{11}$cm$^{-2}$, the resistivity increases with increasing $T$ and is linear above 50 K, suggesting carrier scattering from acoustic phonons. At $T=240$~K the mobility is $\sim120,000$~cm$^2$/Vs, higher than in any known semiconductor. At the charge neutral point we observe a non-universal conductivity that decreases with decreasing $T$, consistent with a density inhomogeneity $<$10$^8$ cm$^{-2}$.
\end{abstract}


\maketitle

Graphene, a single layer of graphite, is a remarkable recent addition to the family of two-dimensional electronic materials. Its linear dispersion relation and the chiral nature of its quasiparticles have created intense experimental and theoretical interest~\cite{rise_graphene}. Central to understanding the electronic transport properties of graphene is the mechanism causing the scattering of its charge carriers. While scanning probe studies show little evidence of intrinsic structural defects in the graphene lattice~\cite{elena,ishigami}, scattering may result from extrinsic sources, such as charged impurities on top of graphene or in the underlying substrate~\cite{fuhrer_charged,dassarma,nomura}, corrugation of the graphene sheet~\cite{geim_intr}, phonons in graphene~\cite{chen_limits,stauber,vasko,dassarma_phonons} or remote interfacial phonons in the substrate~\cite{chen_limits}. The formation of electron and hole puddles can further contribute to scattering at low carrier density~\cite{yacoby,guinea_ripples}.

Recently, dramatically reduced carrier scattering was reported in suspended graphene devices~\cite{us, rutgers}. After an annealing treatment to remove the residual impurities the sample mobility exceeded 200,000~cm$^2$/Vs, an order of magnitude improvement over graphene devices fabricated on a substrate~\cite{us}. The exceptional cleanliness of suspended samples allows us to probe previously inaccessible transport regimes in graphene. In this Letter, we report a strong temperature ($T$) dependence of electrical transport in ultra-clean suspended graphene. At low temperatures, the carrier mean free path in our highest quality devices reaches device dimensions and suggests near-ballistic transport. When the temperature is increased, the resistivity exhibits two distinct behaviors, depending on carrier density. At large densities, the conductivity of graphene exhibits a metallic behavior (i.e., decreasing with increasing $T$) which can be mostly ascribed to electron-phonon scattering. The scattering is remarkably weak, allowing the observation of a very high mobility of $\sim$120,000~ cm$^2$/Vs near-room temperature ($T$=240~ K). At low density, near the charge neutrality point, the conductivity of graphene shows a pronounced non-metallic $T$-dependence (i.e., a decrease with decreasing $T$), indicating a strongly reduced charge inhomogeneity in suspended samples as compared to previously studied unsuspended devices.

\begin{figure}
\includegraphics[width=75mm]{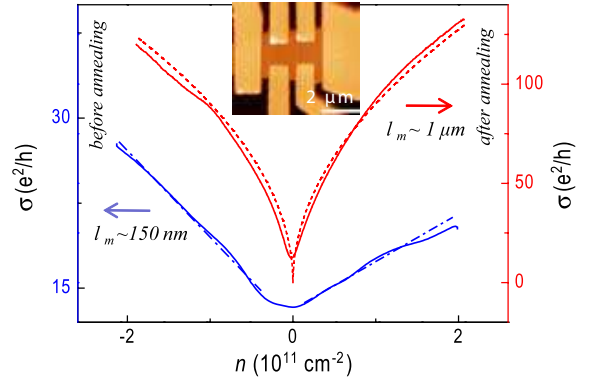}
\caption{(color online) Conductance of the suspended graphene sample S1 before (blue line) and after (red line) annealing as a function of carrier density. Data are shown for $T=40$~K to suppress universal conductance fluctuations. Note the change from near-linear to sub-linear behavior before and after annealing, respectively. The dotted red line is the expectation for ballistic transport (see text). Inset: atomic force microscope image of the suspended device (S1).}
\label{fig1}
\end{figure}

The suspended graphene devices are fabricated using the process described in~\cite{us}. A mechanically-exfoliated graphene flake pressed onto a SiO$_2$/Si substrate is contacted by microlithographically patterned gold electrodes and the SiO$_2$ under the flake is subsequently partially removed via a chemical etch. The fabrication results in flat graphene, suspended $\sim$150~nm above the SiO$_2$/Si substrate, which serves as a gate (Fig.~1, Inset). The electrical measurements are performed in a sample-in-vacuum cryostat capable of $T=$5-240~K. At yet larger $T$ the sample quality can degrade, probably due to a rising background pressure and absorption of impurities onto the graphene. The measurements consist of recording the resistivity $\rho$ as a function of gate voltage $V_g$ and temperature $T$. The gate voltage is limited to $|V_g| <5$~V to avoid electrostatic collapse of the suspended graphene~\cite{us}.  Multiple temperature and voltage sweeps are performed to ensure the reproducibility of the features observed in $\rho(V_g,T)$. The carrier density $n$ is determined via Hall measurements. Assuming a parallel plate capacitor geometry, we find $n=C_g(V_g-V_{NP})/e$, with $C_g=60$ aF/$\mu$m$^2$ and $V_{NP}$ being the gate voltage position of the charge-neutrality point (NP). Since $|V_{NP}|<1$~V and is $T$-independent for all devices, we deduce that the features in $\rho(V_g,T)$ are intrinsic and not caused by the absorption/desorption of impurities.

Before current annealing, the low-$T$ conductivity $\sigma=1/\rho$ of our suspended devices depends linearly on $n$, with mobility $\mu=\sigma/en \sim  28,000$~cm$^2$/Vs (Fig.~1, Sample S1, lower line), comparable to conventional samples fabricated on a substrate. Sending a large current through the device and heating the graphene to an estimated 400 $^\circ$C \cite{us,bachtold} improves the mobility to 170,000~cm$^2$/Vs at $n=2\times10^{11}$ cm$^2$ (Fig.~1, upper line)~\cite{us}, while $\sigma(n)$ becomes nonlinear. A similar improvement is observed for two other suspended devices, sample S2 ($\mu=60,000$~cm$^2/Vs$) and sample S3 ($\mu = 70,000$~cm$^2/Vs$).

We can gain insight into the dominant low-$T$ scattering mechanism in graphene by comparing $\sigma(n)$ before and after current annealing. The linearity of $\sigma(n)$ before annealing suggests the dominance of charged impurity scattering \cite{yanwen, nomura, fuhrer_charged}. For screened Coulomb potential scattering, the scattering time is $\tau \propto k_F$~\cite{nomura,dassarma}  which leads to a conductivity $\sigma=\frac{2e^2}{h}k_F v_F \tau\propto n$, with $k_F$  being the Fermi wavevector. The mean free path in these devices is $l_m = \sigma h/2e^2 k_F \sim 150$~nm, much smaller than the sample size ($> 1~\mu$m), justifying the use of the Boltzmann model for unannealed devices.

In contrast, after current annealing the mean free path in S1 increases to $\sim$ 1 $\mu$m, comparable to the device dimensions of $\sim$2~$\mu$m and transport is no longer diffusive. To elucidate the actual transport conditions in our sample we compare $\sigma(n)$ with the expectation from the extreme opposite position of purely ballistic propagation. There, the current is carried by the finite number of longitudinal modes $N=Wk_F/\pi$, where $W$ is the width of the sample. Assuming perfect transmission, the ballistic conductance at vanishing $T$ is given by~\cite{beenakker,guinea_ballistic}
\begin{equation}
\sigma_{bal}(n) = \frac{4e^2}{h}N = \frac{4e^2}{h}\frac{Wk_F}{\pi} \propto  \sqrt{n}
\end{equation}
The dotted curve in Fig.~1 shows the result of Eq.~(1) for a width $W$=1.3~$\mu$m, close to the dimensions of the device. From the excellent agreement both in shape and   magnitude, combined with the derived long mean free path, we conclude that the low-$T$ transport in our current-annealed, suspended devices is close to the ballistic limit. As an important consequence, the peak mobility appears to be limited by boundary scattering set by the device dimensions and not by impurity scattering and yet larger mobilities should be achievable. Sub-linear behavior of $\sigma(n)$, reminiscent of Fig.~1, was previously observed in unsuspended samples~\cite{yanwen,geim_intr} and interpreted as a combined contribution of short-range and long-range scatterers. Given that the device dimension is comparable to the mean free path, such an interpretation based on the diffusive transport does not seem to be warranted for our samples.

\begin{figure}
\includegraphics[width=75mm]{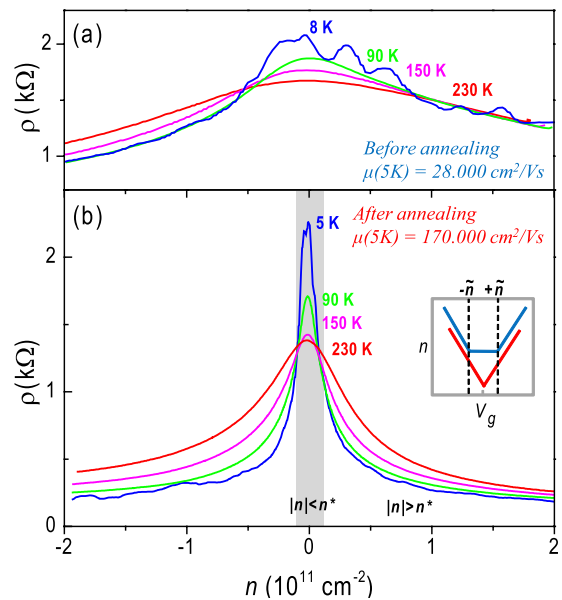}
\caption{(color online) $T$-dependence of resistance of suspended device S1 before (a) and after (b) current annealing. Inset: Sketch of gate voltage dependence of the carrier density in clean (lower curve) and charge inhomogeneous (upper curve) graphene.}
\label{fig2}
\end{figure}

The $T$-dependence of the resistivity provides a tool to investigate the impact of the current annealing process. Before current annealing, the resistivity of device S1 exhibits a relatively small ($< 30\%$) variation of the resistivity from 5~K to 240~K (Fig.~2b), similar to conventional unsuspended devices~\cite{geim_intr,yanwen-ejp}, whereas after annealing this variation is very pronounced ($>200\%$) (Fig.~2a, Fig.~4,Inset). These $T$-dependent data can be divided into two different density regimes, separated by $n^*$, the density at which $\rho(n^{*})$ is $T$-independent (Fig 2b). For $|n|<n^*$ annealed devices exhibit a non-metallic behavior (increasing $\rho$ for decreasing $T$), with the change in peak resistivity as large as a factor of three in the vicinity of NP. For $|n|>n^*$ the resistivity exhibits metallic behavior (decreasing $\rho$ for decreasing $T$). In this regime $\rho(T)$ is generally linear in $T$ above a device specific crossover temperature $T^*$ ($<50$~K), and the slope of $\rho(T)$ increases for smaller $n$ (Fig. 3).

The $T$-dependence observed in unsuspended graphene is considerably different. There, $\rho(T)$ is approximately linear for $T\leq100$~K, while it is superlinear for $T>100$~K~\cite{geim_intr,chen_limits}. The lack of such activated behavior in our suspended devices shows that such activated behavior is not an intrinsic property of graphene, but rather stems from the external sources, such as remote interface phonons~\cite{chen_limits} or static ripples~\cite{geim_intr}. In fact, suspended graphene shows only a modest increase of the resistivity from $T\sim$5~K to 240~K maintaining a mobility of $\mu=120,000$~cm$^2$/Vs at T=240~K and at our highest density of $n=2 \times 10^{11}$~cm$^{-2}$. This value considerably exceeds the highest reported room-temperature mobility for a semiconducting material (InSb, 77,000~cm$^2$/Vs~\cite{insb}).

In order to quantify the linear rise of $\rho(T)$ in Fig.~3 we define the slope in the high-$T$ range as $\Delta \rho/\Delta T=[\rho(200K)-\rho(100K)]/100$~K . The inset to Fig.~3 shows $\Delta \rho/ \Delta T$ as a function of $n$ in samples S1 and S2 before (dotted lines) and after (solid lines) current annealing. Several features stand out. First, the high mobility states reached in S1 and S2 after current annealing exhibit very similar $\Delta \rho/ \Delta T(n)$ dependencies, in spite of a mobility difference of a factor of 2. Second, in the large $n$ limit, the slope $\Delta \rho/\Delta T(n)$ is similar for the samples before and after the annealing process, indicating that the slope in this limit is rather insensitive to the sample mobility. Finally, for all devices $\Delta \rho/\Delta T$ is consistently larger for negative $V_g$ than for positive $V_g$.

\begin{figure}
\includegraphics[width=75mm]{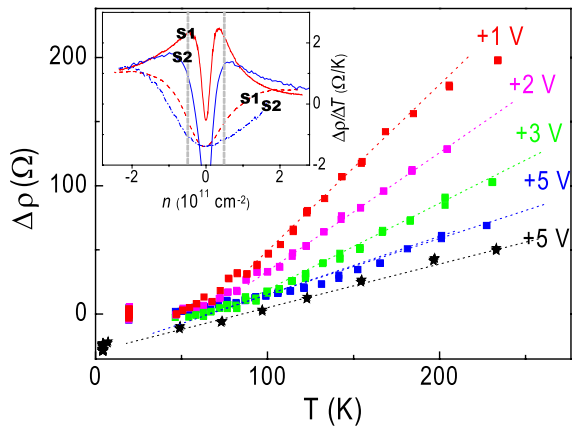}
\caption{(color online) $T$-dependence of the resistivity in sample S2 ($\blacksquare$) and S1 ($\bigstar$, shifted for clarity) at several different gate voltages. Inset: Density dependence of slope of $\Delta \rho/\Delta T$ defined as  $\Delta \rho/ \Delta T=[\rho(200$~K$)-\rho(100$~K$)] /100$~K for sample S1 and S2 before (dotted line) and after (solid line) current annealing. Since $\rho(T)$ is linear only for $n >$ 0.5$\times$10$^{11}$cm$^{-2}$, the definition of  $\Delta \rho/\Delta T$ is only meaningful outside of the dotted region.}
\label{fig3}
\end{figure}

We separately consider the two density regimes, $|n|>n^*$ and $|n|<n^*$. At high densities, the linearly increasing $\rho(T)$ suggests electron-phonon interaction as the dominant source of carrier scattering~\cite{chen_limits,vasko,stauber,dassarma_phonons}. Indeed, within a Boltzmann model and for sufficiently high $T > T_{BG}=2\hbar v_{ph}k_F/k_B \sim 23$~K (BG=Bloch-Gruneisen) at $n=2\times10^{11}$~cm$^{-2}$, the resistivity is linear in $T$
\begin{equation}
\Delta \rho=\frac{\pi D^2 k_B T}{4e^2\hbar \rho_m v_F^2 v_{ph}^2}
\end{equation}
where $D$ is the deformation potential, $\rho_m=7.6\times10^{-8}$~g/cm$^2$ is the graphene mass density, $v_{ph}=2\times10^4$~m/s is the LA phonon velocity~\cite{phonon_parameters} and $v_F=1\times10^6$~m/s is the Fermi velocity~\cite{rise_graphene}. While our data in Fig.~3 show clearly such a linear $T$-dependence, indicative of phonon scattering, the slope of $\rho(T)$  displays an unexpected density dependence, not captured by Eq.~(1). The origin of this density dependence is unclear. It may point to additional contributions from a different $T$-dependent scattering mechanisms at lower densities, such as screened Coulomb scattering~\cite{falko}. Further experimental and theoretical work is needed to resolve this issue. However, for large $|n|$, when $\Delta \rho/\Delta T(n)$ in Fig.~3b reaches a roughly $n$-independent value (at least for positive $V_g$) we may identify this limiting behavior with exclusively phonon scattering and derive an upper bound value for $D$. For $n=+2\times10^{11}$~cm$^{-2}$ (electrons) Eq. (1) yields $D \sim$ 29 eV, consistent with $D=10-30$~eV in graphite \cite{dassarma_phonons, phonon_parameters} and comparable to $D\sim17$~eV, reported for unsuspended graphene \cite{chen_limits}. In contrast, for $n=-2\times10^{11}$~cm$^{-2}$ (holes) we obtain $D\sim50$~eV. This value may be overestimated, since $\Delta\rho/\Delta T (n)$ is not fully saturated even at $V_g \approx $-5~V, the experimental limit of hole density. Nevertheless, this large asymmetry is unusual and presently unresolved and, together with the observed $n$ dependence, may point to a scattering behavior in suspended graphene that is more complex than simple electron-phonon interaction.

We now turn to the low density regime, $|n|<n^*$, and address the $T$-dependence of the minimum conductivity, $\sigma_{min}$. Figure~4 shows $\sigma_{min}(T)/\sigma_{min}(5K)$ in samples S1, S2 and S3 before and after annealing. Before annealing, $\sigma_{min}(T)$ varies only slightly from 5 K to 240 K ($<30\%$). The variation is similar to $\sigma_{min}(T)$ in unsuspended samples of similar mobilities~\cite{yanwen-ejp}. This is in a sharp contrast to the current-annealed devices, where $\sigma_{min}$  acquires a strong $T$-dependence, as large as a factor of $1.5-3$ for $T=5$~K to 250~K. This confirms that the $T$-independence of $\sigma_{min}$ observed in low mobility samples~\cite{yanwen-ejp,geim_intr} is not an intrinsic property of graphene.

\begin{figure}
\includegraphics[width=75mm]{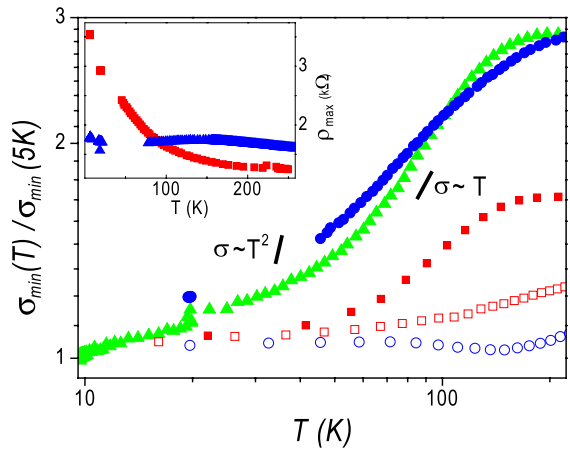}
\caption{(color online) Minimum conductivity normalized to its value at $T=5$~K as a function of $T$ for three devices before (S1:\textcolor{red}{$\square$},S2:\textcolor{blue}{$\circ$}) and after (S1:\textcolor{red}{$\blacksquare$},S2:\textcolor{blue}{$\bullet$},S3:\textcolor{green}{$\blacktriangle$}) annealing.  Data for S3 before current annealing are not available. The $T$-dependence increases considerably after annealing. Inset: Maximum resistivity for S2 before (\textcolor{blue}{$\blacktriangle$}) and after (\textcolor{red}{$\blacksquare$}) current annealing. }
\label{fig4}
\end{figure}

The remarkable property of graphene to exhibit $\sigma_{min}\sim e^2/h$ even at vanishing charge density has been the subject of several experimental and theoretical investigations~ \cite{geim_intr,beenakker,dassarma,dassarma_suspended,fuhrer_charged}. In a high mobility sample, a $T$-dependence is only expected for $k_B T > \epsilon_F = \hbar v_F k_F = \hbar v_F\sqrt{\pi C_g(V_g-V_{NP})/e}$ \cite{dassarma_bilayer,falko}, when $V_g \rightarrow V_{NP}$. However, in a realistic sample charged impurities~ \cite{chen_limits,nomura,dassarma,yacoby} or structural disorder~ \cite{guinea_ripples} break up the carrier system into puddles of electrons and holes for $V_g \rightarrow V_{NP}$.  As a result, the combined (electron plus hole) carrier density in "dirty" graphene never drops below a value  $\tilde{n}$  , referred to as inhomogenity density (Fig.~2, Inset).

Estimating the rms chemical potential in the puddle regime to be $\epsilon_F\sim \hbar v_F \sqrt{\pi\tilde{n}}$ , we expect a significant $T$-dependence of $\sigma_{min}(T)$ only for $k_B T \gg \hbar v_f \sqrt{\pi\tilde{n}}$. For unsuspended samples $\tilde{n}\sim10^{11}$~cm$^{-2}$~\cite{yanwen,dassarma} corresponding to $T_F^*= \hbar v_f \sqrt{\pi\tilde{n}}/k_B \sim 400$~K, which is used to explain why in "dirty" devices   $\sigma_{min}$ lacks a significant $T$-dependence in the range [0, 300~K]. The weak $T$-dependence in suspended samples before current annealing implies a similar situation. In contrast, after current annealing $\sigma_{min}(T)$ shows $T$-dependence down to $T\sim10$~K (Fig.~4) This implies an upper bound of $\tilde{n}<10^8$~cm$^{-2}$, consistent with fits to the density dependence of transport data~\cite{dassarma_suspended} and represents further evidence for the high quality of suspended graphene.

While qualitatively our data suggest that the $T$-dependence of $\sigma_{min}(T)$ in suspended samples results from a low $\tilde{n}$, a quantitative description of $\sigma_{min}$ is complicated by the fact that $k_F l_m>1$ at small $n$ and a Boltzmann description may no longer be applicable. Such complicating factors are evident from a comparison between different annealed devices, which exhibit both a different magnitude and a different functional $T$-dependence as seen in Fig.~4. This suggests that the transport at the neutrality point in the suspended samples is dominated by extrinsic scattering. Interestingly, $\sigma_{min}(T)$ exhibits a $T$-dependence, which is much weaker than the $\sigma_{min}\propto T^2$ (solid line in Fig.~4), expected from a Boltzmann model~\cite{geim_intr,dassarma_bilayer}.

In conclusion, we demonstrate that at low $T$, suspended, current annealed graphene can sustain near-ballistic transport over micron dimensions. At high temperatures, the resistivity of such high-quality graphene increases linearly with $T$. The origin is likely phonon scattering, but a density and carrier-type dependence raises questions as to our present understanding of transport in such devices. The deduced upper bound of the deformation potential, $D$, in the high density limit is comparable to values from graphite, but varies considerably between electron and hole transport.  Finally, the observed strong $T$-dependence of $\sigma_{min}$ in high mobility suspended devices is consistent with very low inhmogeneity density $\tilde{n} < 10^8$~cm$^{-2}$.

We acknowledge experimental help from and discussions with E. Henriksen, M. Foster, S. Adam, I. Aleiner, V. Fal'ko, M. Fuhrer and A. Geim. This work is supported by the NSF (No. DMR-03-52738), NSEC grant CHE-0641523, NYSTAR, DOE (No. DE-AIO2-04ER46133 and No. DEFG02-05ER46215), ONR (No. N000150610138), FENA MARCO, W. M. Keck Foundation, and the Microsoft Project Q.

%\end{thebibliography}

\end{document}